\documentclass[12pt]{article}
\usepackage{amsmath,amssymb,bm,epsfig,color,graphicx}
\renewcommand{\thefootnote}{\fnsymbol{footnote}}

\begin{document}

\title{
\begin{flushright}
\begin{minipage}{0.2\linewidth}
\normalsize
EPHOU-17-02 \\
WU-HEP-17-02\\*[50pt]
\end{minipage}
\end{flushright}
{\Large \bf 
Axion Inflation and Affleck-Dine Baryogenesis
\\*[20pt]}}

\author{
Kensuke Akita$^{1,}$\footnote{
E-mail address: ken8a1@asagi.waseda.jp},\ \ 
Tatsuo~Kobayashi$^{2}$\footnote{
E-mail address:  kobayashi@particle.sci.hokudai.ac.jp},\ and \ 
Hajime~Otsuka$^{1,}$\footnote{
E-mail address: h.otsuka@aoni.waseda.jp
}\\*[20pt]
$^1${\it \normalsize 
Department of Physics, Waseda University, 
Tokyo 169-8555, Japan} \\
$^2${\it \normalsize 
Department of Physics, Hokkaido University, Sapporo 060-0810, Japan} 
\\*[50pt]
}

\date{
\centerline{\small \bf Abstract}
\begin{minipage}{0.9\linewidth}
\medskip 
\medskip 
\small
String theory generically predicts the coupling between the Affleck-Dine field and axion 
field through higher-dimensional operators. 
We thus explore the Affleck-Dine baryogenesis on an axion background. 
It turns out that the axion oscillation produces an enough amount of baryon 
asymmetry of the Universe just after the inflation, even without a soft supersymmetry-breaking $A$-term. 
This baryogenesis scenario is applicable to the string axion inflation. 
\end{minipage}
}

\begin{titlepage}
\maketitle
\thispagestyle{empty}
\end{titlepage}

\renewcommand{\thefootnote}{\arabic{footnote}}
\setcounter{footnote}{0}

\tableofcontents
\clearpage

\section{Introduction}
Axions are attractive fields which are not only able to solve the strong CP problem in the standard 
model, but also promising candidates of dark matter and inflaton field. 
These axion fields are also motivated in the consistent theory of quantum gravity such as 
string theory, since a lot of axions naturally appear in the low-energy effective theory through 
the compactification of extra-dimensional space. 
These string axions are originating from  higher-form fields such as Kalb-Ramond field, 
Ramond-Ramond field, and internal metric. Their gauge symmetries require the axion potential 
to be flat. 
Thus, the string axion has a shift symmetry in the low-energy effective theory. 
Although this continuous shift symmetry is non-perturbatively broken to the discrete one, 
we can control the axion potential with help of the discrete shift symmetry. 
In this regard, the string axion plays a role of inflaton field. 

The string axion inflation models are mainly categorized as follows. 
When the continuous shift symmetry is broken by the non-perturbative effects such as 
the world-sheet instanton, D-brane instanton, and gaugino condensation on hidden D-branes, 
the natural inflation~\cite{Freese:1990rb,Abe:2014pwa} and modulated natural inflation~\cite{Czerny:2014wza,Abe:2014xja,Kappl:2015esy,Choi:2015aem} 
are achieved by employing the 
idea of alignment mechanism~\cite{Kim:2004rp} and loop-enhancement~\cite{Abe:2014pwa}. 
On the other hand, when the axionic shift symmetry is broken by the existence of 
brane, the axion monodromy inflation can be realized~\cite{Silverstein:2008sg,McAllister:2008hb}.\footnote{See also 
Ref.~\cite{Kaloper:2008fb}.} 
In particular, the $F$-term axion monodromy inflation is also a promising candidate, which 
is discussed in Refs.~\cite{Marchesano:2014mla,Blumenhagen:2014gta} and derived in type IIB string theory with three-form 
fluxes~\cite{Hebecker:2015rya,Kobayashi:2015aaa}. 
In any axion inflation model, the baryon asymmetry of the Universe has to be sufficiently produced in order 
to be consistent with the Big-Bang Nucleosynthesis  and cosmic microwave background data. 

In this paper, we focus on the baryogenesis scenario after the axion inflation with an emphasis 
on the Affleck-Dine (AD) baryogenesis scenario~\cite{Affleck:1984fy,Dine:1995kz}. 
The AD baryogenesis scenario is interesting, since it is applicable even when there are dilution 
effects induced by the moduli oscillation. 
In the conventional AD baryogenesis, the Hubble-induced $A$-term and soft supersymmetry (SUSY) breaking 
$A$-term are taken as the complex constants. The phase differences between them are then important 
to obtain an enough amount of baryon asymmetry. 
However, in superstring theory as well as higher-dimensional theory, these $A$-terms are 
moduli- and axion-dependent in general. 
These $A$-terms, i.e., the Yukawa and higher-dimensional couplings, 
are obtained through the integration of matter wavefunction over  
extra-dimensional space, and then  obtained couplings depend on the size and 
shape of extra dimensions, which correspond to the moduli and axion fields from the four-dimensional 
point of view. 
Indeed, in type IIB string theory on a toroidal background with magnetized D$7$-branes, 
$n$-point couplings among the matter fields on magnetized branes depend on 
the axion associated with the complex structure moduli at the perturbative level~\cite{Cremades:2004wa,Abe:2009dr}. 
Furthermore, non-perturbative effects such as the brane-instanton 
also generate the potential for both the axion and matter fields, where the axion is 
identified with the K\"ahler axion associated with the internal cycle of Calabi-Yau (CY) 
manifold~\cite{Blumenhagen:2006xt,Ibanez:2006da,Florea:2006si}. 
Similarly, Yukawa couplings and higher dimensional couplings are also calculated, 
e.g. in heterotic string theory on orbifolds \cite{Hamidi:1986vh,Dixon:1986qv,Burwick:1990tu,Erler:1992gt,Choi:2007nb} 
and type IIA intersecting D-brane models \cite{Cvetic:2003ch,Abel:2003vv,Cremades:2003qj,Abel:2003yx}.
These include world-sheet instanton effects, and 
the couplings are moduli- and axion-dependent.

When the axion plays a role of inflaton field, the dynamics of the axion field would change that of 
the AD field during and after the inflation. 
We thus explore the AD baryogenesis on this axion-inflaton background.\footnote{In type IIB string theory 
on Calabi-Yau manifold, AD baryogenesis is considered in the context of LARGE volume scenario~\cite{Higaki:2012ba,Allahverdi:2016yws}.}
The remaining of this paper is organized as follows. We first briefly review the conventional AD baryogenesis in Sec.~\ref{sec:2_1}. 
By outlining the higher-order couplings among the matter fields in Sec.~\ref{sec:2_2}, 
we study the dynamics of the AD field on an axion background after the inflation in Sec.~\ref{sec:2_3} 
and during the inflation in Sec.~\ref{sec:2_4}. 
Finally, Sec.~\ref{sec:con} is devoted to the conclusion. 
In Appendix~\ref{app}, we discuss the explicit model which combines both the axion inflation 
and SUSY-breaking sectors.

\section{Affleck-Dine baryogenesis on an axion background}
\label{sec:2}
\subsection{Conventional AD baryogenesis}
\label{sec:2_1}
First of all, we briefly review the conventional AD baryogenesis. 
In the supersymmetric standard model, the scalar potential at the renormalizable level vanishes 
along the flat directions composed by squarks and sleptons 
in the global limit. (For details of flat directions, 
see, e.g., Ref.~\cite{Dine:1995kz}.) 
These flat directions are parametrized by the AD field. 
When the AD field ($\Phi$) has a baryon charge $\beta$, the baryon asymmetry can be 
generated by the dynamics of the AD field, 
\begin{align}
&n_B=i\beta \left( \frac{d\Phi^\ast}{dt} \Phi -\Phi^\ast \frac{d\Phi}{dt}\right)=2\beta \phi^2 \frac{d\theta}{dt},
\end{align}
where $\Phi \equiv\phi e^{i\theta}$ and $t$ is the cosmic time. 
It implies that when radial and phase directions of the AD field have nonvanishing 
vacuum expectation value (VEV) and velocity, the baryon asymmetry can be produced. 

The scalar potential of the AD field is lifted by the soft SUSY-breaking and non-renormalizable 
operators. In particular, the non-renormalizable superpotential term, which is invariant under the gauge 
symmetries of the standard model and R-parity, can be written  as follows,\footnote{AD baryogenesis with 
R-parity violating operators is discussed in Ref.~\cite{Higaki:2014eda}.}
\begin{align}
W=\lambda \frac{\Phi^n}{M_{\rm Pl}^{n-3}},
\label{eq:WAD}
\end{align}
where $\lambda$ is the complex constant and $n=4,5,\cdots,$ depending on the choice of 
flat direction in the supersymmetric standard model. 
By taking into account the SUSY-breaking effects, the relevant Lagrangian density of 
the AD field is described in the following form,
\begin{align}
\sqrt{-g}{\cal L}&=a^3\left[-\frac{1}{2}\partial_\mu \Phi \partial^\mu \Phi^\ast -V\right],
\end{align}
where $g$ is the determinant of Friedmann-Robertson-Walker metric $g_{\mu\nu}={\rm diag}(1,-a^2,-a^2,-a^2)$ with $\mu,\nu=0,1,2,3$ 
and $a$ being the scale factor. The total scalar potential of the AD field $V$ is described by
\begin{align}
V&=(m_{\Phi}^2-cH^2)|\Phi|^2 +|\lambda|^2\frac{|\Phi|^{2(n-1)}}{M_{\rm Pl}^{2n-6}}
-\left(\frac{a_H\lambda H\Phi^n}{M_{\rm Pl}^{n-3}}+\rm{h.c.}\right)
-\left(\frac{a_m\lambda m_{3/2}\Phi^n}{M_{\rm Pl}^{n-3}}+\rm{h.c.}\right),
\label{eq:pocon}
\end{align}
where $c$ and $a_{H,m}$ are the constants, $m_{\Phi}$, $m_{3/2}$ and $H=d\ln a/dt$ are the 
soft-mass of the AD field, gravitino mass and Hubble-parameter, respectively. 
The two terms including $H$ correspond to the Hubble-induced mass term and Hubble-induced $A$-term.
As long as $m_{\Phi},m_{3/2}<H$ during and after the inflation, the radial direction of the AD field $\phi$ becomes 
a nonvanishing VEV, $\phi_{\rm min}^{(n)}\simeq \left(\frac{c^{1/2}HM_{\rm Pl}^{n-3}}{\sqrt{n-1}|\lambda|}\right)^{\frac{1}{n-2}}$, 
while the phase direction of the AD field $\theta$ is fixed at 
$n\theta \simeq -{\rm arg}(a_H)-{\rm arg}(\lambda)$ modulo $2\pi \mathbb{Z}$ because of the Hubble-induced $A$-term. 
When the Hubble parameter decreases to the mass of $\phi$, i.e., $H\simeq m_{\Phi}$, 
the AD field begins to oscillate around the minimum at the time $t_{\rm osc}\simeq m_{\Phi}^{-1}$. 
The phase direction of the AD field $\theta$ is kicked by the difference between ${\rm arg}(a_H)$ and 
${\rm arg}(a_m)$. The baryon asymmetry is then produced and after this time $t_{\rm osc}$, 
the baryon to entropy ratio is fixed. 
Eventually, we obtain a sizable amount of baryon asymmetry of the Universe. 
Along the above procedure, the phase difference between Hubble and soft SUSY-breaking 
$A$-terms is important to a nonvanishing velocity of $\theta$. 
Note that the baryon asymmetry can be also produced even without the Hubble-induced $A$-term, 
since the quantum fluctuation of $\theta$ during and after the inflation, $m_{\Phi},m_{3/2}<H$, 
triggers the oscillation of the AD field to the direction of $\theta$ at $t\simeq t_{\rm osc}$. 
However, in this case, the sizable isocurvature perturbation of $\theta$ is severely constrained 
by the Planck data~\cite{Planck:2013jfk} as discussed in Refs.~\cite{Enqvist:1998pf,Enqvist:1999hv,Kawasaki:2001in,Kasuya:2008xp}. 

So far, we have treated the $n$-point coupling $\lambda$ and $A$-terms as the complex constants. 
However, these are generically moduli- and axion-dependent on the basis of string theory. 
Especially, when the axion field plays a role of inflaton field, the dynamics of axion-inflaton would change 
that of the AD field. 
Following this line of thought, 
we aim to examine the dynamics of the AD field with an emphasis on the 
coupling between the AD field and the axion-inflaton.

\subsection{The origin of higher-order coupling in string theory}
\label{sec:2_2}
Before going to the detail of the AD scenario on an axion background, 
we mention about the origin of higher-order couplings in string theory. 

At the perturbative level, $n$-point couplings for matter fields are calculated in the 
type IIB string theory on the toroidal background, in particular, three factorizable tori $(T^2)^3$. 
For example, in Ref.~\cite{Cremades:2004wa}, the Yukawa couplings among the matter fields are derived in 
the field theoretical approach. When we consider the matter fields  on 
magnetized D-branes, these wavefunctions are quasi-localized in the extra-dimensional space 
because of the magnetic fluxes $M_i$ on each torus $(T^2)_i$, $i=1,2,3$.
Indeed, by computing the overlap integral among the matter fields, 
the three-point couplings on two-torus $T^2$ are described by the Jacobi theta-function $\vartheta$,
\begin{align}
y^{(3)}_{ijk}\simeq \sum_{m\in {\bm Z}_{M_3}}~\delta_{i+j+M_1m,k} \vartheta 
\begin{bmatrix}
\frac{M_2i-M_1j+M_1M_2m}{M_1M_2M_3}\\
0
\end{bmatrix}
(0, i\tau M_1M_2M_3) \propto e^{i\varphi/f},
\end{align}
up to a normalization factor, 
where ${\bm Z}_{M_3}={1,2,\cdots, |M_3|}$ with $M_3=M_1+M_2$. Here, $\varphi$ is a 
canonically normalized axion associated with the complex structure modulus of torus $\tau$, 
and $f$ is its decay constant.(See 
for the detail, Refs.~\cite{Cremades:2004wa,Abe:2009dr}.) 
Furthermore, in Ref.~\cite{Abe:2009dr}, this analysis is extended to the higher-order couplings for 
the matter fields. The $n$-point couplings are written by the product of three-point couplings 
through the relation,
\begin{align}
y_{i_1,i_2,\cdots,i_n}^{(n)}=\sum_s y_{i_1,i_2,\cdots,i_{n-2}s}^{(n-1)}\cdot  y^{(3)}_{\bar{s}i_{n-1}i_n}.
\end{align}
That leads to the relation, $y^{(n)} \sim (y^{(3)})^{n-2}$.
We remark that the magnitude of $n$-point couplings is typically smaller than unity, since it 
is determined by the product of Yukawa couplings among the fields in the standard model. 
Also, in the T-dual intersecting D-brane models, Yukawa couplings and higher dimensional couplings 
are obtained through world-sheet instanton effects in stringy calculation~\cite{Cvetic:2003ch,Abel:2003vv,Cremades:2003qj,Abel:2003yx} and 
they depend on the K\"ahler moduli as well as their axionic parts.
Their stringy calculations are similar to calculations of Yukawa couplings and higher dimensional couplings 
in heterotic string theory on orbifolds \cite{Hamidi:1986vh,Dixon:1986qv,Burwick:1990tu,Erler:1992gt,Choi:2007nb}.
Moreover, in heterotic string theory on CY, the holomorphic Yukawa couplings among the matter fields also depend on 
the axion associated with the complex structure moduli~\cite{Strominger:1985ks,Candelas:1987se} in a similar fashion.

Next, we briefly comment on $n$-point couplings at the non-perturbative level through the 
world-sheet instanton and brane-instanton effects~\cite{Blumenhagen:2006xt,Ibanez:2006da,Florea:2006si}. 
For example,  in type IIB string theory, Euclidean brane instanton such as E$3$-instanton 
generates the non-perturbative potential for the K\"ahler modulus $T$, i.e., $e^{-2\pi T}$, 
where $T$ is the modulus chiral superfield associated with the divisor in CY. 
However, when another D-brane intersects with this Euclidean brane, the charged open 
string modes propagating in these D-branes $\Phi_i$ with $i=1,2,\cdots,n$ 
appear in the low-energy effective theory. 
Thus, the gauge invariance requires the superpotential in the following form,
\begin{align}
W\propto e^{-2\pi T}\Phi_1\Phi_2\cdots \Phi_n.
\end{align}
Then, the $n$-point coupling among the matter fields is a function of the axion 
corresponding to the imaginary part of the K\"ahler modulus. 
In such a case, $n$-point couplings are also suppressed by the volume of internal cycle such that  
 the supergravity approximation is valid, $T\gg 1$. 
However, since the value of $\lambda$ is correlated with the energy density of $\Phi$ through 
$V_{\rm AD}\simeq -cH_{\rm inf}^2|\Phi|^2\simeq -cH_{\rm inf}^2(\phi_{\rm min}^{(4)})^2$ during the inflation, 
$\lambda$ is bounded below by a requirement that the energy density of the AD field does not dominate 
that of Universe, i.e., $V_{\rm AD}<V_{\rm inf}\simeq 3H_{\rm inf}^2M_{\rm Pl}^2$, e.g., for $n=4$, 
\begin{align}
|\lambda|>8\times 10^{-8}c^{3/2}\left(\frac{H_{\rm inf}}{10^{12}\,{\rm GeV}}\right),
\end{align}
and for $n=6$, 
\begin{align}
|\lambda|>2.1\times 10^{-8}c^{5/2}\left(\frac{H_{\rm inf}}{10^{12}\,{\rm GeV}}\right).
\end{align}

\subsection{Dynamics of the AD field}
\label{sec:2_3}
To provide a concrete analysis, we take the following ansatz of $n$-point coupling of the AD fields,
\begin{align}
\lambda =|\lambda|e^{i\varphi/f},
\end{align}
where $\varphi$ is the axion-inflaton with $f$ being its decay constant. Here, we assume that the other closed and open 
string moduli are heavier than the axion-inflaton and then they are set as the vacuum expectation values, 
otherwise other moduli fields would destroy the successful axion inflation. 
This assumption is justified for the specific axion inflation model discussed in Appendix~\ref{app}. 
The absolute value of moduli-dependent coupling $|\lambda|$ is thus set as the constant. 
By taking into account the SUSY-breaking effects, the potential of the AD field is described as
\begin{align}
V&=(m_{\Phi}^2-cH^2)|\Phi|^2 +|\lambda|^2\frac{|\Phi|^{2(n-1)}}{M_{\rm Pl}^{2n-6}}
\nonumber\\
&-\frac{2|a_H| |\lambda| H|\Phi|^n}{M_{\rm Pl}^{n-3}}\cos \left( n\theta +{\rm arg}(a_H)+\frac{\varphi}{f}\right)
-\frac{2|a_m| |\lambda| m_{3/2}|\Phi|^n}{M_{\rm Pl}^{n-3}}\cos \left( n\theta +{\rm arg}(a_m)+\frac{\varphi}{f}\right),
\label{eq:po}
\end{align}
with $a_{H(m)}=|a_{H(m)}|e^{i{\rm arg}(a_{H(m)})}$.
Then, as long as $m_{\Phi},m_{3/2}\ll H$ during and after the inflation, 
the minimum of the AD field, in particular, $\theta$ depends on the axion-inflaton, 
\begin{align}
\phi_{\rm min}^{(n)} &\simeq \left(\frac{c^{1/2}HM_{\rm Pl}^{n-3}}{\sqrt{n-1}|\lambda|}\right)^{\frac{1}{n-2}}
\simeq \left(\frac{\alpha^{(n)} M_{\rm Pl}^{n-3}}{t|\lambda|}\right)^{\frac{1}{n-2}},
\nonumber\\
n\theta_{\rm min} &\simeq -\frac{\varphi}{f}-{\rm arg}(a_H)+2l\pi,
\label{eq:thetamin}
\end{align}
where $\alpha^{(n)}=\sqrt{4c/9(n-1)}$ and $l\in \mathbb{Z}$. 
Here and in what follows, the energy density of the Universe is 
assumed to be dominated by that of inflaton, i.e., $H\simeq 2/(3t)$. 
When the axion-inflaton begins to oscillate around the minimum, the axion potential is well described in 
the quadratic form, $V_{\rm inf}\simeq \frac{1}{2}m_{\rm inf}^2\varphi^2$, where $m_{\rm  inf}$ is the mass of 
the axion-inflaton. 
In this era, the axion-inflaton behaves as
\begin{align}
\varphi(t)\simeq \sqrt{\frac{8}{3}}\frac{M_{\rm Pl}}{m_{\rm inf}t} \sin(m_{\rm inf}t).
\label{eq:infmo}
\end{align}
Therefore, the time-dependent axion motion would lead to the nonvanishing velocity of $\theta$. 
It implies that the axion oscillation gives rise to the baryon asymmetry of the Universe in 
this epoch. 

To justify our statement, we first analyze the dynamics of the AD field with $n=4$ in Sec.~\ref{sec:2_3_1} 
and $n=6$ in Sec.~\ref{sec:2_3_2} after the inflation. 
The dynamics of the AD field during the inflation will be discussed in Sec.~\ref{sec:2_4}, where we specify 
the axion-inflation scenario.

\subsubsection{Model $1$ ($n=4$)}
\label{sec:2_3_1}
In this section, we study the dynamics of the AD field which couples with the axion through the $4$-point coupling $\lambda$. 
First of all, we analytically solve the dynamics of the AD field after the inflation, $m_{3/2}, m_{\Phi} < H<H_{\rm inf}$. 
In this era, the axion-inflaton oscillates around the minimum and it behaves as in Eq.~(\ref{eq:infmo}). 
On this axion background, we solve the equations of motion for the AD field, $\chi =|\Phi|(\phi_{\rm min}^{(n=4)})^{-1}$ and $\theta={\rm arg}(\Phi)$, where 
$ \phi_{\rm min}^{(n=4)}\simeq \left(\frac{\alpha^{(n=4)} M_{\rm Pl}}{t|\lambda|}\right)^{1/2}$
is the minimum of radial direction of the AD field 
after the inflation, $m_{3/2}, m_{\Phi} < H$. 
In the axion oscillating era, $m_{3/2}, m_{\Phi} < H$, the scalar potential is approximately given by 
\begin{align}
V&\simeq -cH^2|\Phi|^2 +|\lambda|^2\frac{|\Phi|^{2(n-1)}}{M_{\rm Pl}^{2n-6}}
-\frac{2|a_H| |\lambda| H|\Phi|^n}{M_{\rm Pl}^{n-3}}\cos \left( n\theta +{\rm arg}(a_H)+\frac{\varphi}{f}\right),
\label{eq:poninf}
\end{align}
and then the equations of motion for $\chi$ and $\theta$ reduce to be
\begin{align}
\ddot{\chi}-\left(\frac{8}{9}c +\frac{1}{4}\right)\chi-(\dot{\theta})^2\chi +\frac{8}{9}c \chi^5
-\frac{16}{3}|a_H|\alpha^{(n=4)} \chi^3\cos(n\theta-n\theta_{\rm min})
=0,
\nonumber\\
\ddot{\theta}+2\frac{\dot{\chi}}{\chi}\dot{\theta}
+\frac{16}{3}|a_H|\alpha^{(n=4)} \chi^2 \sin(n\theta -n\theta_{\rm min})=0. 
\label{eq:EOM4}
\end{align}
Throughout this paper, the dot denotes the derivative with respect to a variable $z=\ln (m_{\rm inf}t)$.

Since there is no damping term for $\chi$ in Eq.~(\ref{eq:EOM4}), $\chi$ just oscillates 
around the minimum $\chi_{\rm min}$. 
To solve the dynamics of $\theta$ analytically, we assume that $\chi$ settles into the minimum $\chi_{\rm min}$ 
and on this hypersurface $\chi=\chi_{\rm min}$, the equation of motion of $\theta$ is 
approximately given by
\begin{align}
\ddot{\theta}+\frac{16}{3}|a_H|\alpha^{(n=4)} \sin(n\theta -n\theta_{\rm min})=0.
\end{align}
Furthermore, in the regime $z\gg1$ ($t\gg m_{\rm inf}^{-1})$, 
$\theta_{\rm min}$ is negligible. 
Therefore, the equation of motion of $\theta$ can be rewritten as
\begin{align}
\ddot{\theta}+\frac{16}{3}|a_H| \alpha^{(n=4)}  \sin(n\theta )=0.
\end{align}
This equation can be solved in terms of an elliptic function ${\rm sn}(z)$, 
\begin{align}
{\rm sn}^{-1}\left[\sin(2\theta(z))\right]-{\rm sn}^{-1}\left[\sin(2\theta(z=0))\right] 
&=\left(\pm 2\sqrt{E+\frac{16}{3n}|a_H|\alpha^{(n=4)}}\right) z,
\end{align}
with
\begin{align}
E=\frac{1}{2}\left(\frac{d\theta}{dz}\right)^2\biggl|_{z=0} -\frac{4}{3}|a_H|\alpha^{(n=4)} \cos(n\theta(z=0)).
\end{align}
As a result, $\theta$ has a nonvanishing velocity 
\begin{align}
\frac{d\theta}{dz} &=\sqrt{2E+\frac{8}{3}|a_H|\alpha^{(n=4)} \cos(n\theta)},
\label{dthetadz}
\end{align}
which becomes the constant. Then, $d\theta/dt$ decreases with the inverse of cosmic time $t$. 
The nonvanishing velocity of $\theta$ gives rise to the baryon asymmetry of the Universe 
in the regime $m_{3/2}, m_{\Phi} < H$, 
\begin{align}
n_B=2\beta |\Phi|^2 \frac{d\theta}{dt}
&\simeq 2\beta \chi_{\rm min}^2(\phi_{\rm min}^{(n=4)})^2t^{-1}\sqrt{2E+\frac{8}{3}|a_H|\alpha^{(n=4)} \cos(n\theta)}
\nonumber\\
&=2\beta \chi_{\rm min}^2\frac{\alpha^{(n=4)} M_{\rm Pl}}{|\lambda|}
t^{-2}\sqrt{2E+\frac{8}{3}|a_H|\alpha^{(n=4)} \cos(n\theta)}.
\label{eq:nBosc4}
\end{align}

Next, let us solve the dynamics of the AD field after the oscillating time of $\phi$, i.e., $t_{\rm osc}\simeq m_{\Phi}^{-1}$. 
During $m_{\Phi},m_{3/2}>H$, the scalar potential is approximately given by
\begin{align}
V&\simeq m_{\Phi}^2(\phi)^2 -\frac{|a_m| |\lambda| m_{3/2}\phi^4}{M_{\rm Pl}}\cos \left( 4\theta +{\rm arg}(a_m)+\frac{\varphi}{f}\right),
\label{eq:pon4osc}
\end{align}
and the equation of motion of $\phi$ reduces in the regime $\phi \ll 1$, 
\begin{align}
&\frac{d^2\phi}{dt^2}+3H\frac{d\phi}{dt} -\phi\left(\frac{d\theta}{dt}\right)^2+\frac{\partial V}{\partial \phi}
\simeq \frac{d^2\phi}{dt^2}+\frac{2}{t}\frac{d\phi}{dt} +2m_\Phi^2  \phi
=0,
\label{eq:eqn4phiosc}
\end{align}
with $H=2/(3t)$. Here, we further employ the following approximation
\begin{align}
m_{\Phi}\gg \left|\frac{d\theta}{dt}\right| =t^{-1}\left|\frac{d\theta}{dz}\right|,
\label{eq:apploxi24}
\end{align}
with $d\theta/dz$ in Eq.~(\ref{dthetadz}). 
This approximation is valid after the oscillation of the AD field, i.e., $t_{\rm reh}>t>t_{\rm osc}\simeq m_{\Phi}^{-1}$, 
unless we take the large initial velocity of $\theta$ at the end of inflation.
(We can confirm this assumption in our numerical calculation.) 
Thus, we can solve the equation of motion of $\phi$ in Eq.~(\ref{eq:eqn4phiosc}) 
by employing the virial theorem. The solution behaves as 
\begin{align}
\phi \simeq \left(\frac{c^{1/2}m_{\Phi}M_{\rm Pl}}{\sqrt{3}|\lambda|}\right)^{1/2}\frac{\sin(m_{\Phi}t)}{m_{\Phi}t},
\end{align}
where we take $\phi(t=t_{\rm osc})\simeq \left(\frac{c^{1/2}m_{\Phi}M_{\rm Pl}}{\sqrt{3}|\lambda|}\right)^{1/2}$.

Let us next take a closer look at the equation of motion of $\theta$,
\begin{align}
&\frac{d^2\theta}{dt^2}+3H\frac{d\theta}{dt}+2\phi^{-1}\frac{d\phi}{dt}\frac{d\theta}{dt} +\phi^{-2}\frac{\partial V}{\partial \theta}
\nonumber\\
&\simeq 
\frac{d^2\theta}{dt^2}+2m_{\Phi}{\rm cot}(m_{\Phi}t)\frac{d\theta}{dt} 
+8\frac{|a_m| |\lambda| m_{3/2}\phi^2}{M_{\rm Pl}}\sin \left( 4\theta +{\rm arg}(a_m)+\frac{\varphi}{f}\right)
\nonumber\\
&\simeq 
\frac{d^2\theta}{dt^2}+2m_{\Phi}{\rm cot}(m_{\Phi}t)\frac{d\theta}{dt} =0,
\end{align}
which is valid in the regime $\phi \ll 1$. 
Thus, it turns out that the velocity of $\theta$ is 
considered to be constant during $t_{\rm osc}<t<t_{\rm reh}$ with $t_{\rm reh}$ being the time at the reheating,
\begin{align}
\frac{d\theta}{dt}\biggl|_{t=t_{\rm reh}} \simeq \left(\frac{\sin (m_{\Phi}t_{\rm reh})}{\sin (m_{\Phi}t_{\rm osc})}\right)^{-2}
\frac{d\theta}{dt}\biggl|_{t=t_{\rm osc}}
\simeq \frac{d\theta}{dt}\biggl|_{t=t_{\rm osc}}.
\end{align}

When the phases of $A$-terms, ${\rm arg}(a_H)$ and ${\rm arg}(a_m)$, are different from each other, 
$d\theta/dt$ at $t\simeq t_{\rm osc}$ depends on them in a way similar to the conventional AD baryogenesis. 
However, in this paper, we assume ${\rm arg}(a_H)={\rm arg}(a_m)$ in contrast to the conventional one 
as discussed in Appendix~\ref{app}, in which 
we determine the phases of the $A$-terms by taking into account the explicit moduli stabilization with the axion inflation. 
We stress that even in this case ${\rm arg}(a_H)={\rm arg}(a_m)$, the baryon asymmetry of the Universe 
is sufficiently produced as shown in Eq.~(\ref{eq:nBosc4}). 

Along the above procedure, we find 
\begin{align}
a^3 n_B=2\beta a^3 \phi^2 \frac{d\theta}{dt}={\rm const.},
\end{align}
during $t_{\rm osc}<t<t_{\rm reh}$ and the baryon to entropy ratio is fixed at $t\simeq t_{\rm osc}$. 
As a result, the baryon to entropy ratio at the reheating temperature is obtained by employing the 
result of Eq.~(\ref{eq:nBosc4}),
\begin{align}
\frac{n_B}{s}&=\frac{1}{s(t_{\rm reh})}
\left(\frac{a(t_{\rm osc})}{a(t_{\rm reh})}\right)^3n_B(t_{\rm osc})
\simeq 
\frac{9}{16}\frac{T_{\rm reh}}{m_\phi^2M_{\rm Pl}^2}
n_B(t_{\rm osc})
\nonumber\\
&\simeq \frac{9}{8}
\frac{T_{\rm reh}}{|\lambda|M_{\rm Pl}}
\alpha^{(n=4)}\beta \chi_{\rm min}^2 
{\rm max}\left\{m_{\rm inf}^{-1}\frac{d\theta}{dt}\biggl|_{t=t_{\rm end}}, \sqrt{\frac{8\alpha^{(n=4)} |a_H|}{3}  \bigl|\cos(n\theta(t_{\rm osc}))-
\cos(n\theta(t_{\rm end}))}\bigl|\right\}.
\label{eq:nBsn41}
\end{align}
When the velocity of $\theta$ is much larger than unity in the unit of inflaton mass, 
the baryon asymmetry is proportional to the velocity of $\theta$ at the end of inflation,
\begin{align}
\frac{n_B}{s}&\simeq 1.8\times 10^{-10}
\left(\frac{T_{\rm reh}}{10^{6}\,{\rm GeV}}\right)\left(\frac{10^{-3}}{|\lambda|}\right)
\beta c^{1/2}\chi_{\rm min}^2 
\times \left( m_{\rm inf}^{-1}\frac{d\theta}{dt}\biggl|_{t=t_{\rm end}}\right).
\label{eq:nBsn42}
\end{align}
As discussed in Sec.~\ref{sec:2_2}, the $n$-point coupling $\lambda$ is typically much smaller than 
unity in the string setup. 
This small $\lambda$ not only enhances the amount of baryon asymmetry but also 
suppresses the thermal corrections for the AD field~\cite{Dine:1995kz,Allahverdi:2000zd,Asaka:2000nb}.

\begin{figure}[htbp]
 \begin{minipage}{0.5\hsize}
  \begin{center}
    \includegraphics[width=7.0cm]{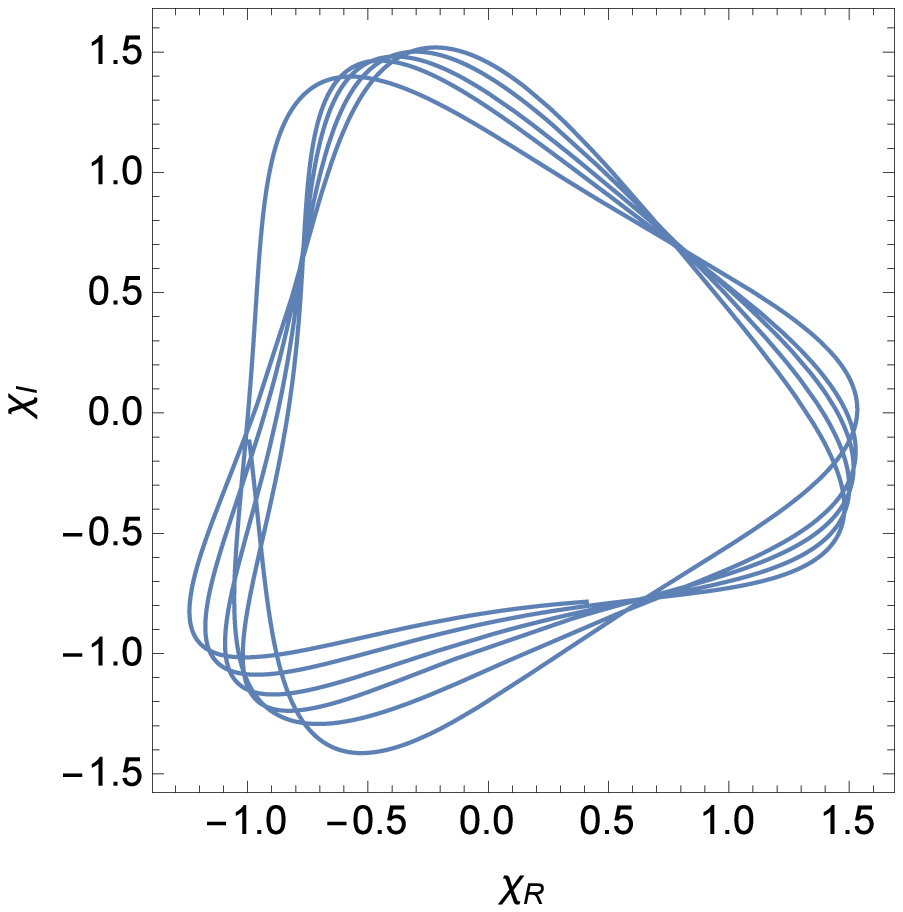}
  \end{center}
 \end{minipage}
 \begin{minipage}{0.5\hsize}
  \begin{center}
   \includegraphics[width=70mm]{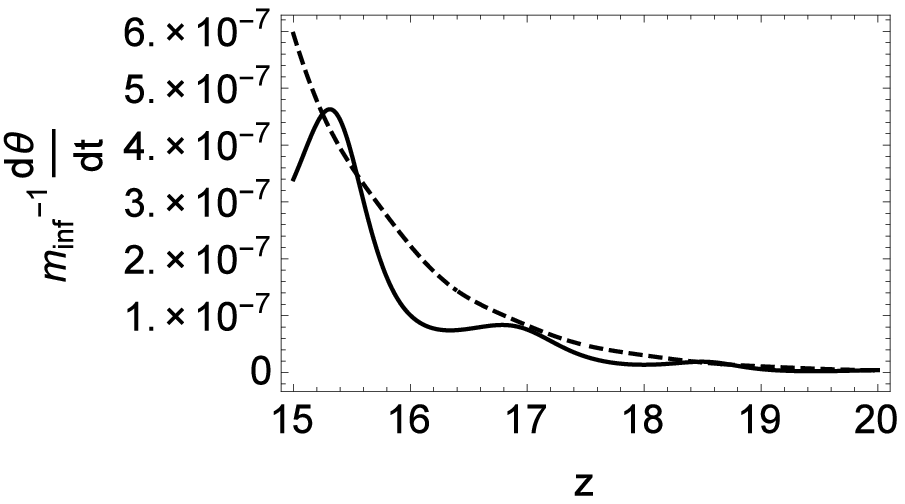}
  \end{center}
 \end{minipage}
     \caption{The dynamics of the AD field for $n=4$. In the left panel, we draw 
     the trajectories of ($\chi_R$,$\chi_I$) as a function of $z=\ln (m_{\rm inf}t)$ with $m_{\rm inf}=10^{12}[\rm GeV]$, whereas, in the right panel, the black solid (dotted) curve corresponds to the numerical (analytical) solution of 
     $m_{\rm inf}^{-1}|d\theta/dt|$ 
     with respect to the same $z$. 
     In both panels, we set the same initial conditions for $(\chi_R, \chi_I)$ satisfying Eq.~(\ref{eq:n4ini2}) 
     and the parameters given in Eq.~(\ref{eq:n4para}).}
         \label{fig:n41}
\end{figure}

To conform our analytical solution, we numerically solve the dynamics of the AD field after inflation. 
In Fig.~\ref{fig:n41}, we draw the trajectories of $(\chi_R={\rm Re}\,\chi, \chi_I={\rm Im}\,\chi)$ and 
$m_{\rm inf}^{-1}\frac{d\theta}{dt}$ 
as a function of $z$ by setting the illustrative parameters: 
\begin{align}
|\lambda|=10^{-3},\qquad 
c=\frac{9}{4},\qquad
\beta=1,\qquad
|a_H|=\frac{1}{8},\qquad
f=4\times 10^{15}[{\rm GeV}], 
\label{eq:n4para}
\end{align}
and the initial conditions of the AD field:
\begin{align}
|\chi|\bigl|_{t=t_{\rm end}}=1,\qquad \theta\bigl|_{t=t_{\rm end}}=\theta_{\rm min}\bigl|_{t=t_{\rm end}},\qquad \frac{d|\chi|}{dt}\biggl|_{t=t_{\rm end}}=0,\qquad 
m_{\rm inf}^{-1}\frac{d\theta}{dt}\biggl|_{t=t_{\rm end}}=\frac{d\theta}{dz}\biggl|_{t=t_{\rm end}}=2,
\label{eq:n4ini}
\end{align}
which correspond to the initial conditions of ($\chi_R,\chi_I$) as
\begin{align}
&\chi_R=\cos(-250\sin(1))\simeq -0.99,\qquad 
\chi_I=\sin(-250\sin(1))\simeq -0.11, \nonumber\\
&\dot{\chi_R}=2\sin(-250\sin(1))\simeq  -0.24,\qquad 
\dot{\chi_I}=-2\cos(-250\sin(1))\simeq 1.99. 
\label{eq:n4ini2}
\end{align}
As shown in the right panel in Fig.~\ref{fig:n41}, the numerical and analytical solutions are well consistent in 
the limit of $z\gg 1$, where $z\simeq 16(21)$ corresponds to the low(high)-scale SUSY breaking $m_{\Phi}\simeq 10^{3}(10^5)$GeV. Although we now adopt the nonvanishing velocity of $\theta$ at the end of inflation, 
we find that $\theta$ has a sizable velocity even when the initial velocity of the AD field becomes zero. 

\begin{figure}[htbp]
 \begin{minipage}{0.5\hsize}
  \begin{center}
    \includegraphics[clip,width=7.0cm]{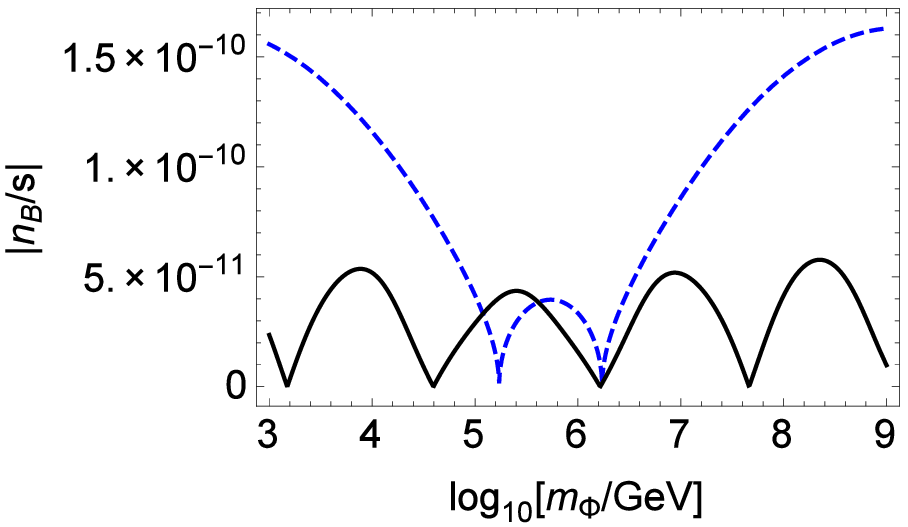}
   \end{center}
  \end{minipage}
 \begin{minipage}{0.5\hsize}
  \begin{center}
   \includegraphics[width=70mm]{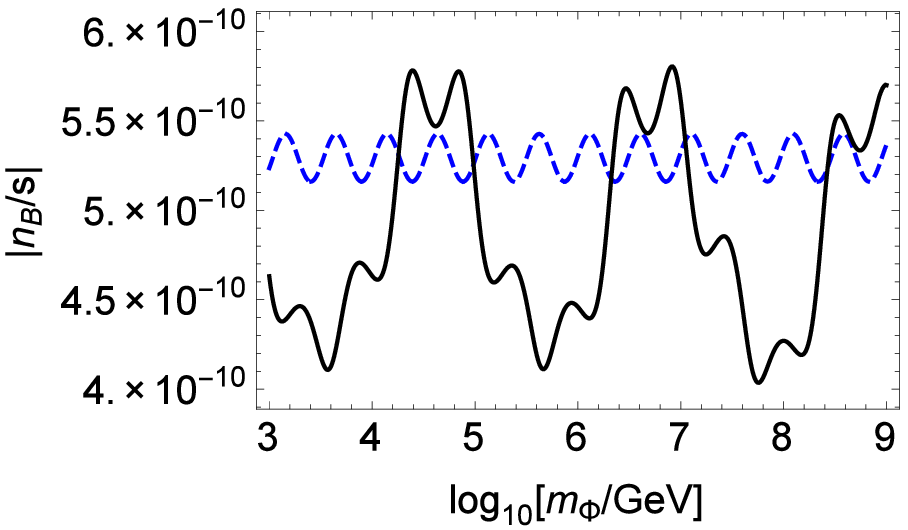}
  \end{center}
 \end{minipage}
        \caption{The baryon to entropy ratio $n_B/s$ at the reheating $T_{\rm reh}=10^6[\rm GeV]$ 
        as a function of mass of the AD field $m_{\Phi}[{\rm GeV}]$. 
   In both panels, the black solid and blue dotted curves represent the numerical and analytical solutions with $\chi_{\rm min}=1$ in Eq.~(\ref{eq:nBsn41}) by setting the same parameters as in Fig.~\ref{fig:n41}. 
    In the left panel, we set the initial conditions at the end of inflation as 
    $(|\chi|,\theta)=(\chi_{\rm min},\theta_{\rm min})$ and $(\dot{|\chi|}, \dot{\theta})=(0,0)$, 
    whereas, in the right panel, the initial conditions for the AD field are the same as in Fig.~\ref{fig:n41}.}
        \label{fig:n42}
\end{figure}

\begin{figure}[htbp]
  \begin{center}
    \includegraphics[clip,width=7.0cm]{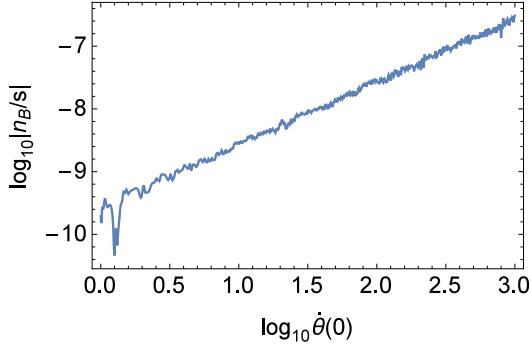}
  \end{center}
    \caption{The baryon to entropy ratio $n_B/s$ at the reheating $T_{\rm reh}=10^6[\rm GeV]$ as a function of $\dot{\theta}(0)=m_{\rm inf}^{-1}\frac{d\theta}{dt}\bigl|_{t=t_{\rm end}}$.  The blue curve corresponds to the numerical solution within the range $1\leq \dot{\theta}(0) \leq 10^3$ by setting 
    $m_{\Phi}=10^3[{\rm GeV}]$, and the other parameters and initial conditions are the same as in Fig.~\ref{fig:n41}.}
            \label{fig:n43}
\end{figure}

In Fig.~\ref{fig:n42}, we show the baryon to entropy ratio $n_B/s$ at the reheating $T_{\rm reh}=10^6[\rm GeV]$ as a function of mass of the AD field $m_\Phi[\rm GeV]$ 
by setting the same parameters in Fig.~\ref{fig:n41}. 
It turns out that the analytical formula of baryon asymmetry in Eq.~(\ref{eq:nBsn41}) 
is well consistent with the numerical one without depending on the value of initial velocity of $\theta$ and 
the mass of the AD field. 
In particular, as shown in the left panel in Fig.~\ref{fig:n42}, the baryon asymmetry is sufficiently produced even 
when the initial velocity of $\theta$ is set to be zero. 
Furthermore, Fig~\ref{fig:n43} also shows the numerical estimation of the baryon to entropy ratio $n_B/s$ at the reheating $T_{\rm reh}=10^6[\rm GeV]$ as a function of $\dot{\theta}(0)=m_{\rm inf}^{-1}\frac{d\theta}{dt}\bigl|_{t=t_{\rm end}}$ 
by setting $m_{\Phi}=10^3[{\rm GeV}]$ and the same parameters and initial conditions in Fig.~\ref{fig:n41}. 
This figure indicates that $n_B/s$ is proportional to the velocity of $\theta$ when $\dot{\theta}(0)\gg 1$ as confirmed in the analytical one in Eq.~(\ref{eq:nBsn42}).

From the analytical estimation of the baryon asymmetry, 
the baryon asymmetry is independent of the decay constant of the axion-inflaton. 
However, the velocity of $\theta$ at the end of inflation is sensitive to this decay constant 
in our numerical calculation. 
We will come back this point in Sec.~\ref{sec:2_4} where the dynamics of the AD field is 
studied during the axion inflation.

So far, when the AD field oscillates around the minimum at $t\simeq t_{\rm osc}$, 
we have assumed that the energy density of the Universe is dominated by the oscillation of 
the axion-inflaton. 
This assumption is valid only if the reheating temperature satisfies the following inequality,
\begin{align}
T_{\rm reh}<\left(\frac{90}{\pi^2 g_{\ast}}\right)^{1/4}\sqrt{m_{\Phi}M_{\rm Pl}}
\simeq 2.2\times 10^{11}\,{\rm GeV}\left(\frac{m_\Phi}{10^5\,{\rm GeV}}\right)^{1/2},
\label{eq:trehu}
\end{align}
where $g_{\ast}\simeq 915/4$ is the effective degree of freedom at the reheating for the 
minimal supersymmetric standard model. 
Furthermore, oscillating time of the AD field should be earlier than the time at the reheating 
$t_{\rm reh}\simeq 1/\Gamma_{\rm dec}$ with the total decay width of inflaton $\Gamma_{\rm dec}$, 
i.e., $t_{\rm osc}<t_{\rm reh}$. 
This situation can be achieved under
\begin{align}
10^{-3}\left(\frac{10^3\,{\rm GeV}}{m_{\Phi}}\right)
< 2.4\times 10^6 \left(\frac{90}{\pi^2 g_{\ast}}\right)^{-1/2}\left(\frac{10^6\,{\rm GeV}}{T_{\rm reh}}\right)^2,
\label{eq:timeu}
\end{align}
and our illustrative parameters satisfy the above conditions in Eqs.~(\ref{eq:trehu}) and (\ref{eq:timeu}).

\vspace{5cm}
\subsubsection{Model $2$ ($n=6$)}
\label{sec:2_3_2}
In this section, we study the dynamics of the AD field which couples with the axion through the $6$-point coupling $\lambda$. 
In the same way with Sec.~\ref{sec:2_3_1}, we first 
analytically solve the dynamics of the AD field after the inflation, $m_{3/2}, m_{\Phi} < H<H_{\rm inf}$. 
In this era, the axion-inflaton oscillates around the minimum and it behaves as in Eq.~(\ref{eq:infmo}). 
On this axion background, we solve the equations of motion for the AD field, $\chi =|\Phi|(\phi_{\rm min}^{(n=6)})^{-1}$ and $\theta={\rm arg}(\Phi)$, where 
$\phi_{\rm min}^{(n=6)}\simeq \left(\frac{\alpha^{(n=6)} M_{\rm Pl}^3}{t|\lambda|}\right)^{1/4}$ is the minimum of radial direction of the AD field 
after the inflation, $m_{3/2}, m_{\Phi} < H$. 
In the axion oscillating era, $m_{3/2}, m_{\Phi} < H$, the equations of motion for $\chi$ and $\theta$ 
are given in terms of a variable $z=\ln (m_{\rm inf}t)$,
\begin{align}
\ddot{\chi}+\frac{1}{2}\dot{\chi}-\left(\frac{8}{9}c +\frac{3}{16}\right)\chi
-(\dot{\theta})^2\chi +\frac{8c}{9} \chi^9-8|a_H|\alpha^{(n=6)} \chi^5\cos(6\theta-6\theta_{\rm min})=0,
\nonumber\\
\ddot{\theta}+\frac{\dot{\theta}}{2}+2\frac{\dot{\chi}}{\chi}\dot{\theta}
+8|a_H|\alpha^{(n=6)} \chi^4 \sin(6\theta -6\theta_{\rm min})=0.
\label{eq:eqm6}
\end{align}
Then, $\chi$ and $\theta$ are damped toward their minima. 
Around the minimum $\theta_{\rm min}$, we expand the solution as 
\begin{align}
\theta &=\theta_{\rm min}+\delta \theta,
\end{align}
and the equation of motion of $\chi$ is then simplified at the linear order of $\delta \theta$, 
\begin{align}
\frac{d^2\chi}{dz^2}+\frac{1}{2}\frac{d\chi}{dz}
-\chi\left( \frac{8}{9}c +\frac{3}{16}-\frac{8c}{9} \chi^8 +8|a_H|\alpha^{(n=6)} \chi^4\right)=0.
\label{eq:eomchi}
\end{align}
Here, we drop the term $(\dot{\theta})^2\chi$, since 
$\dot{\theta}$ decreases with time as confirmed later. 
From Eq.~(\ref{eq:eomchi}), $\chi$ has the fixed point:
\begin{align}
\chi_{\ast}^4=\frac{72|a_H|\alpha^{(n=6)}+\sqrt{5184(|a_H|\alpha^{(n=6)})^2+2c(27+128c)}}{16c}.
\end{align}

At the fixed point $\chi=\chi_{\ast}$, $\theta$ obeys the following equation around the minimum $\theta_{\rm min}$,
\begin{align}
\ddot{\theta}+\frac{\dot{\theta}}{2}
+48|a_H|\alpha^{(n=6)} \chi_{\ast}^4(\theta -\theta_{\rm min})=0,
\end{align}
where there is no damping term $\dot{\chi}/\chi$. 
Furthermore, by omitting the term $\theta_{\rm min}\propto e^{-z}$ in the regime $z\gg1$,
the equation of motion of $\theta$ is rewritten
\begin{align}
\ddot{\theta}+\frac{\dot{\theta}}{2}
+48|a_H|\alpha^{(n=6)} \chi_{\ast}^4 \theta=0,
\end{align}
and it can be solved as
\begin{align}
\theta (z)=e^{-z/4}\left[ \theta(0) {\rm cosh} (pz/4)+ p^{-1}\left(\theta(0)+4\dot{\theta}(0)\right){\rm \sinh} (pz/4)\right],
\end{align}
with $p=\sqrt{\frac{5-512|a_H|\chi_{\rm min}^4\sqrt{5c}}{5}}$. 
It turns out that $\theta$ has a nonvanishing velocity in this era, $t_{\rm end}<t<t_{\rm osc}$,
\begin{align}
\frac{d\theta}{dt}=t^{-1}\frac{d\theta}{dz}=
m_{\rm inf}^{-1/4}t^{-5/4}
\left[ \left(\frac{\theta(0)+4\dot{\theta}(0)}{4}\right) {\rm cosh} (pz/4)+ \frac{p\dot{\theta}(0)}{4}{\rm \sinh} (pz/4)\right],
\end{align}
and the produced baryon asymmetry is estimated as
\begin{align}
n_B&=2\beta \phi^2 \frac{d\theta}{dt}\simeq 
2\beta (\phi_{\rm min}^{(n=6)}\chi(t))^2 \frac{d\theta}{dt}
\nonumber\\
&\simeq 2\beta \left(\frac{\alpha^{(n=6)} M_{\rm Pl}^3}{|\lambda|}\right)^{1/2}
m_{\rm inf}^{-1/4}t^{-7/4}
\chi_{\ast}^2
\left[ \left(\frac{\theta(0)+4\dot{\theta}(0)}{4}\right) {\rm cosh} (pz/4)+ \frac{p\dot{\theta}(0)}{4}{\rm \sinh} (pz/4)\right] .
\label{eq:nBosc6}
\end{align}

Next, we focus on the dynamics of the AD field after the oscillation of the AD field, i.e., $t_{\rm osc}<t<t_{\rm reh}$. 
In a way similar  to the discussion in Sec.~\ref{sec:2_3_1}, 
$\phi$ decreases with the inverse of cosmic time $t$, and 
$\theta$ just oscillates around the minimum. 
The velocity of $\theta$ is also determined at $t\simeq t_{\rm osc}$, 
since the equation of motion of $\theta$ after the time $t_{\rm osc}$ is almost the same as in the case of $n=4$ in Sec.~\ref{sec:2_3_1}. 

As mentioned before, we now assume ${\rm arg}(a_H)={\rm arg}(a_m)$ in contrast to the conventional AD scenario. 
In Appendix~\ref{app}, this situation is derived in the explicit moduli stabilization scenario compatible with the axion inflation. 
We stress that even in this case ${\rm arg}(a_H)={\rm arg}(a_m)$, the baryon asymmetry of the Universe 
is sufficiently produced as shown in Eq.~(\ref{eq:nBosc6}). 
Along the above procedure, we find 
\begin{align}
a^3 n_B=2\beta a^3 \phi^2 \frac{d\theta}{dt}={\rm const.},
\end{align}
during $t_{\rm osc}<t<t_{\rm reh}$ and the baryon to entropy ratio is fixed at $t\simeq t_{\rm osc}$. 
We conclude that the baryon to entropy ratio at the reheating temperature is determined by the 
result of Eq.~(\ref{eq:nBosc6}),
\begin{align}
\frac{n_B}{s}\biggl|_{\rm reh}&=\frac{1}{s(t_{\rm reh})}
\left(\frac{a(t_{\rm osc})}{a(t_{\rm reh})}\right)^3n_B(t_{\rm osc})
\simeq
\frac{9}{16}\frac{T_{\rm reh}}{m_\phi^2M_{\rm Pl}^2}n_B(t_{\rm osc})
\nonumber\\
&\simeq 
\frac{9}{8}\beta \frac{T_{\rm reh}}{M_{\rm Pl}^{1/2}}\left(\frac{\alpha^{(n=6)}}{|\lambda|}\right)^{1/2}
m_{\rm inf}^{-1/4}m_{\Phi}^{-1/4}
\chi_{\ast}^2 
\left[ \left(\frac{\theta(0)+4\dot{\theta}(0)}{4}\right) {\rm cosh} (pz_{\rm osc}/4)+ \frac{p\dot{\theta}(0)}{4}{\rm \sinh} (pz_{\rm osc}/4)\right]
\nonumber\\
&\simeq 
7.1\times 10^{-10}\beta c^{1/4}\left(\frac{T_{\rm reh}}{10^2\,{\rm GeV}}\right)
\left(\frac{10^{-5}}{|\lambda|}\right)^{1/2}
\left(\frac{10^{12}\,{\rm GeV}}{m_{\rm inf}}\right)^{-1/4}
\left(\frac{10^{5}\,{\rm GeV}}{m_\Phi}\right)^{-1/4}
\chi_{\ast}^2 
\nonumber\\
&\times
\left[ \left(\frac{\theta(0)+4\dot{\theta}(0)}{4}\right) {\rm cosh} (pz_{\rm osc}/4)+ \frac{p\dot{\theta}(0)}{4}{\rm \sinh} (pz_{\rm osc}/4)\right].
\label{eq:nBsn61}
\end{align}

\begin{figure}[htbp]
 \begin{minipage}{0.5\hsize}
  \begin{center}
    \includegraphics[width=7.0cm]{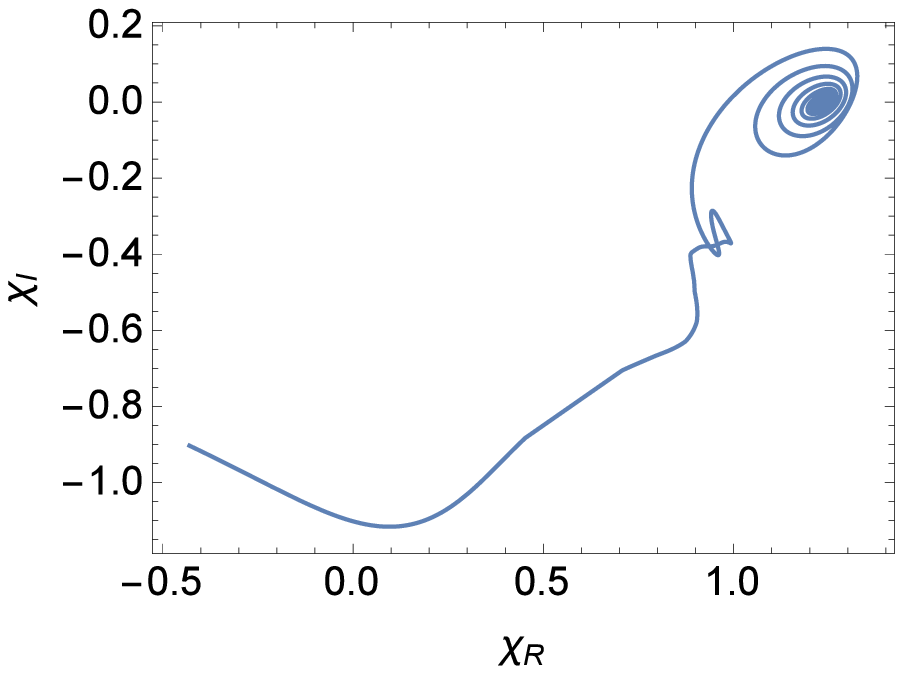}
  \end{center}
 \end{minipage}
 \begin{minipage}{0.5\hsize}
  \begin{center}
   \includegraphics[width=70mm]{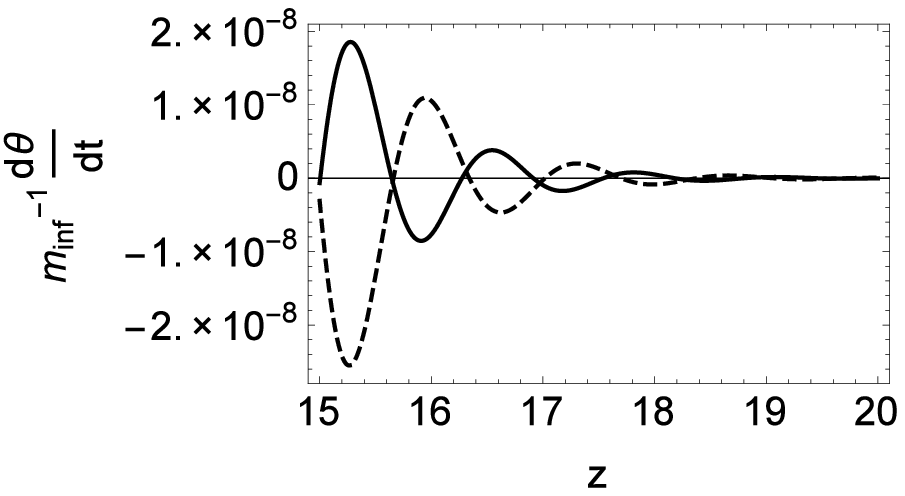}
  \end{center}
 \end{minipage}
     \caption{The dynamics of the AD field for $n=6$. In the left panel, we draw 
     the trajectories of ($\chi_R$,$\chi_I$) as a function of $z=\ln (m_{\rm inf}t)$ with $m_{\rm inf}=10^{12}[\rm GeV]$, 
     whereas,  in the right panel, the black solid (dotted) curve corresponds to the numerical (analytical) solution of 
     $m_{\rm inf}^{-1}|d\theta/dt|$ 
     with respect to the same $z$. 
     In both panels, we set the same initial conditions for $(\chi_R, \chi_I)$ satisfying Eq.~(\ref{eq:n6ini2}) 
     and the parameters given in Eq.~(\ref{eq:n6para}).}
              \label{fig:n61}
\end{figure}

Finally, we numerically solve the equation of motion of the AD field given by Eq.~(\ref{eq:eqm6}). 
In Fig.~\ref{fig:n61}, we draw the trajectories of $(\chi_R={\rm Re}\,\chi, \chi_I={\rm Im}\,\chi)$ and 
$m_{\rm inf}^{-1}\frac{d\theta}{dt}$ 
as a function of $z=\ln (m_{\rm inf}t)$ with $m_{\rm inf}=10^{12}[{\rm GeV}]$ by setting the illustrative parameters: 
\begin{align}
|\lambda|=10^{-3},\qquad 
c=\frac{9}{4},\qquad 
\beta=1,\qquad 
|a_H|=1,\qquad 
f=4\times 10^{15}[{\rm GeV}], 
\label{eq:n6para}
\end{align}
and the initial conditions of the AD field:
\begin{align}
|\chi|\bigl|_{t=t_{\rm end}}=1,\qquad \theta\bigl|_{t=t_{\rm end}}=\theta_{\rm min}\bigl|_{t=t_{\rm end}},\qquad \frac{d|\chi|}{dt}\biggl|_{t=t_{\rm end}}=0,\qquad 
m_{\rm inf}^{-1}\frac{d\theta}{dt}\biggl|_{t=t_{\rm end}}=\frac{d\theta}{dz}\biggl|_{t=t_{\rm end}}=2,
\label{eq:n6ini}
\end{align}
which correspond to the initial conditions of ($\chi_R,\chi_I$) as
\begin{align}
&\chi_R=\cos\left(-\frac{1000}{6}\sin(1)\right)\simeq -0.43,\qquad 
\chi_I=\sin\left(-\frac{1000}{6}\sin(1)\right)\simeq -0.9, \nonumber\\
&\dot{\chi_R}=2\sin\left(-\frac{1000}{6}\sin(1)\right)\simeq  -1.81,\qquad 
\dot{\chi_I}=-2\cos\left(-\frac{1000}{6}\sin(1)\right)\simeq -0.86. 
\label{eq:n6ini2}
\end{align}
As shown in the right panel in Fig.~\ref{fig:n61}, the numerical and analytical solutions are well consistent in 
the limit of $z\gg 1$, where $z\simeq 16(21)$ corresponds to the low(high)-scale SUSY breaking $m_{\Phi}\simeq 10^{3}(10^5)$GeV. Although we now adopt the nonvanishing velocity of $\theta$ at the end of inflation, 
we find that $\theta$ has a sizable velocity even when the initial velocity of the AD field becomes zero. 

\begin{figure}[htbp]
 \begin{minipage}{0.5\hsize}
  \begin{center}
    \includegraphics[clip,width=7.0cm]{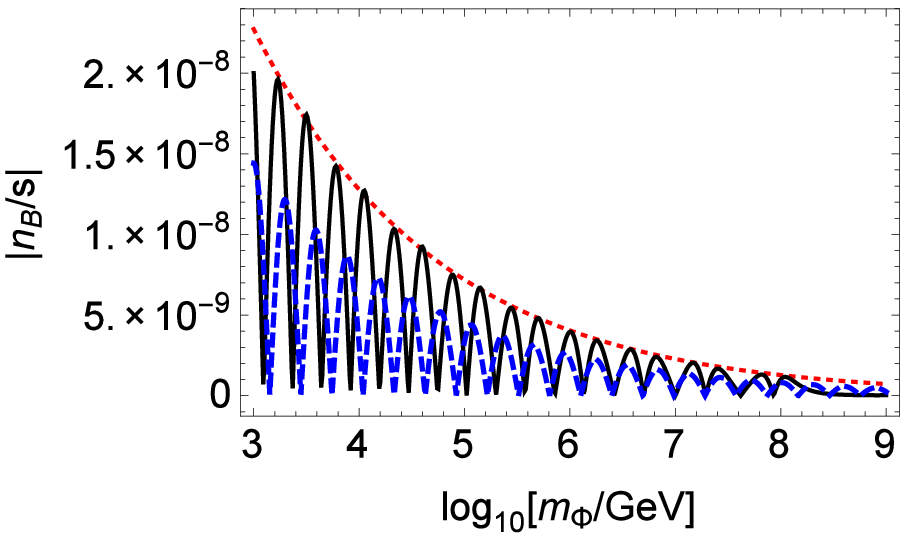}
   \end{center}
	\label{fig:ns_mphi_n=6}
  \end{minipage}
 \begin{minipage}{0.5\hsize}
  \begin{center}
   \includegraphics[width=70mm]{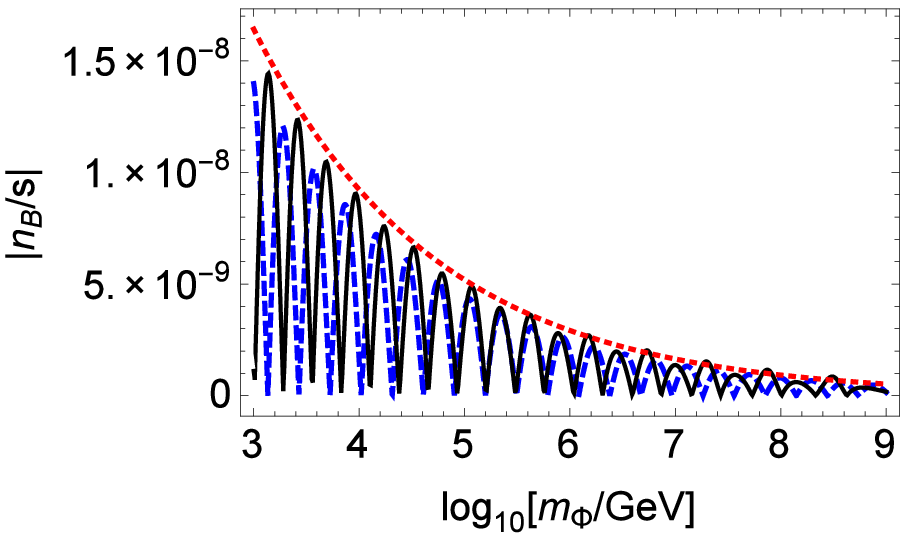}
  \end{center}
    \label{fig:ns_mphi_n=6_v}
 \end{minipage}
\caption{The baryon to entropy ratio $n_B/s$ at the reheating $T_{\rm reh}=10^2[\rm GeV]$ 
        as a function of mass of the AD field $m_{\Phi}[{\rm GeV}]$. 
   In both panels, the black solid and blue dotted curves represent the numerical and analytical solutions with $\chi_{\rm min}=1$ in Eq.~(\ref{eq:nBsn61}) by setting the same parameters as in Fig.~\ref{fig:n61}. The damping of baryon asymmetry is well fitted with the red curves, 
   $1.3\times 10^{-7-\frac{1}{4}\ln_{10}(m_{\Phi}/{\rm GeV})}$ in the left panel and $9.2\times10^{-8-\frac{1}{4}\ln_{10}(m_{\Phi}/{\rm GeV})}$ 
   in the right panel, as expected in the analytical formula~(\ref{eq:nBsn61}). 
    In the left panel, we set the initial conditions at the end of inflation as 
    $(|\chi|,\theta)=(\chi_{\rm min},\theta_{\rm min})$ and $(\dot{|\chi|}, \dot{\theta})=(0,0)$, 
    whereas, in the right panel, the initial conditions for the AD field are the same as in Fig.~\ref{fig:n61}.}
        \label{fig:n62}
\end{figure}

\begin{figure}[htbp]
  \begin{center}
   \includegraphics[width=70mm]{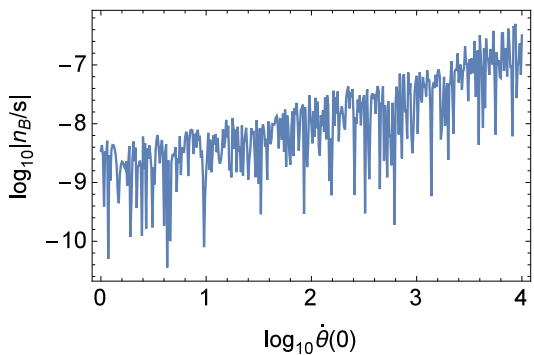}
  \end{center}
    \caption{The baryon to entropy ratio $n_B/s$ at the reheating $T_{\rm reh}=10^2[\rm GeV]$ as a function of $\dot{\theta}(0)=m_{\rm inf}^{-1}\frac{d\theta}{dt}\bigl|_{t=t_{\rm end}}$.  The blue curve corresponds to the numerical solution within the range $1\leq \dot{\theta}(0) \leq 10^4$ by setting 
    $m_{\Phi}=10^5[{\rm GeV}]$, and the other parameters and initial conditions are the same as in Fig.~\ref{fig:n41}.}
            \label{fig:n63}
\end{figure}

With the same parameters in Fig.~\ref{fig:n61}, we show the baryon to entropy ratio $n_B/s$ at the reheating $T_{\rm reh}=10^2[\rm GeV]$ as a function of mass of the AD field $m_\Phi[\rm GeV]$ in Fig.~\ref{fig:n62}. 
It turns out that the analytical formula of baryon asymmetry in Eq.~(\ref{eq:nBsn61}) 
is well consistent with the numerical one without depending on the value of initial velocity of $\theta$. 
In particular, as shown in the left panel in Fig.~\ref{fig:n62}, the baryon asymmetry is sufficiently produced even 
when the initial velocity of $\theta$ is set to be zero. 
Furthermore, Fig~\ref{fig:n63} also shows the numerical estimation of the baryon to entropy ratio $n_B/s$ at the reheating $T_{\rm reh}=10^2[\rm GeV]$ as a function of $\dot{\theta}(0)=m_{\rm inf}^{-1}\frac{d\theta}{dt}\bigl|_{t=t_{\rm end}}$ 
by setting $m_{\Phi}=10^5[{\rm GeV}]$ and the same parameters and initial conditions in Fig.~\ref{fig:n61}. 
This figure indicates that $n_B/s$ is proportional to the velocity of $\theta$ when $\dot{\theta}(0)\gg 1$ as confirmed in the analytical one in Eq.~(\ref{eq:nBsn61}).

In a way similar to Sec.~\ref{sec:2_2}, the analytical formula of the baryon asymmetry is independent of the decay constant of  axion-inflaton. 
However, the velocity of $\theta$ at the end of inflation is sensitive to this decay constant 
in our numerical calculation. 
Thus, we explore the dynamics of the AD field during the inflation in Sec.~\ref{sec:2_4}.

\vspace{10cm}

\subsection{Axion inflation}
\label{sec:2_4}
The results of previous sections~\ref{sec:2_2} and~\ref{sec:2_3} indicate that 
the obtained baryon asymmetry depends on the velocity of $\theta$ just after the inflation. 
To estimate the dynamics of the AD field, we first consider the axion monodromy inflation~\cite{Silverstein:2008sg,McAllister:2008hb}. 
The scalar potential of canonically normalized axion-inflaton $\varphi$ is
\begin{align}
V=V_{\rm AD}(\varphi, \Phi)+V_{\rm inf}(\varphi),
\end{align}
where $V_{\rm AD}$ is the scalar potential of the AD field in Eq.~(\ref{eq:poninf}) and 
\begin{align}
V_{\rm inf}(\varphi)=\Lambda \varphi^q.
\label{eq:infpo2_4}
\end{align}
Here, $\Lambda$ is the real constant and $q$ is the rational number depending on the model~\cite{McAllister:2014mpa}. 
By solving the equation of motion for the inflaton field in the slow-roll regime, 
\begin{align}
\frac{d^2\varphi}{dt^2}+3H\frac{d\varphi}{dt}+\frac{\partial V}{\partial \varphi}
\simeq 3H\frac{d\varphi}{dt}+\frac{\partial V_{\rm inf}}{\partial \varphi}=0,
\end{align}
the axion-inflaton behaves as
\begin{align}
\varphi(t)^{-\frac{q}{2}+2}\simeq \varphi(t=0)^{-\frac{q}{2}+2} -q\sqrt{\frac{\Lambda}{3}}t.
\end{align}
As discussed in Ref.~\cite{Dine:1995kz}, the radial direction of the AD field $\phi$ quickly settles into 
the minimum, 
$\phi_{\rm min}^{(n)} \simeq \left(\frac{c^{1/2}HM_{\rm Pl}^{n-3}}{\sqrt{n-1}|\lambda|}\right)^{\frac{1}{n-2}}$. 
By contrast, the minimum of $\theta$ given by Eq.~(\ref{eq:thetamin}) 
gradually decreases with time, when the decay constant $f$ is larger than the order of Planck scale. 
In such a case, the velocity of $\theta$ is roughly estimated as 
\begin{align}
\frac{d\theta}{dt}\simeq \frac{d\theta_{\rm min}}{dt}\simeq \frac{1}{nf}\frac{d\varphi}{dt}
\simeq -\frac{q}{nf}\sqrt{\frac{\Lambda}{3}}\varphi^{\frac{q}{2}-1}.
\label{eq:anavelo}
\end{align}
On the other hand, for $f<{\cal O}(M_{\rm Pl})$, $\theta$ would not move along the minimum $\theta_{\rm min}$, 
since $\theta_{\rm min}$ quickly moves.

To analyze the dynamics of $\theta$ during the inflation, we thus numerically solve the following equation of motion 
on the hypersurface $\phi=\phi_{\rm min}^{(n)}$,
\begin{align}
&\frac{d^2\theta}{dt^2}+3H\frac{d\theta}{dt}+2(\phi_{\rm min}^{(n)})^{-1}\frac{d\phi_{\rm min}^{(n)}}{dt}\frac{d\theta}{dt} +(\phi_{\rm min}^{(n)})^{-2}\frac{\partial V}{\partial \theta}=0,
\end{align}
with an emphasis on $q=2$ in Eq.~(\ref{eq:infpo2_4}) and $n=4,6$. 
Note that the above equation is independent of $\lambda$. 
Although the cosmological observables predicted in the axion inflation with $q=2$ is disfavored by the Planck data, 
we focus on this case for our illustrative purposes. 
Indeed, the behavior of the AD field for the axion monodromy inflation with other $q$ is similar to the $q=2$ case. 
We thus set the parameters in the scalar potential in Eq.~(\ref{eq:infpo2_4}) such that the power spectrum of 
curvature perturbation $P_\xi$ and its spectral tilt $n_s$ are consistent with the Planck data,
\begin{align}
P_\xi= \frac{V}{24\pi^2M_{\rm Pl}^4\epsilon}\simeq 2.2\times 10^{-9},\qquad n_s=1+2\eta-6\epsilon\simeq 0.96, 
\end{align}
where $\epsilon$ and $\eta$ are the slow-roll parameters:
\begin{align}
\epsilon=\frac{M_{\rm Pl}^2}{2}\left(\frac{\partial_\varphi V}{V}\right)^2,
\qquad
\eta=M_{\rm Pl}^2\left(\frac{\partial_\varphi\partial_\varphi V}{V}\right).
\end{align} 
Then, we find that $H\simeq \sqrt{\Lambda}\left(14-\frac{2\sqrt{\Lambda}t}{\sqrt{3}}\right)$ with $\Lambda=2.4\times 10^{-11}$ in the $M_{\rm Pl}=2.4\times 10^{18}\,[{\rm GeV}]=1$ unit.

\begin{figure}[htbp]
 \begin{minipage}{0.5\hsize}
  \begin{center}
    \includegraphics[clip,width=7.0cm]{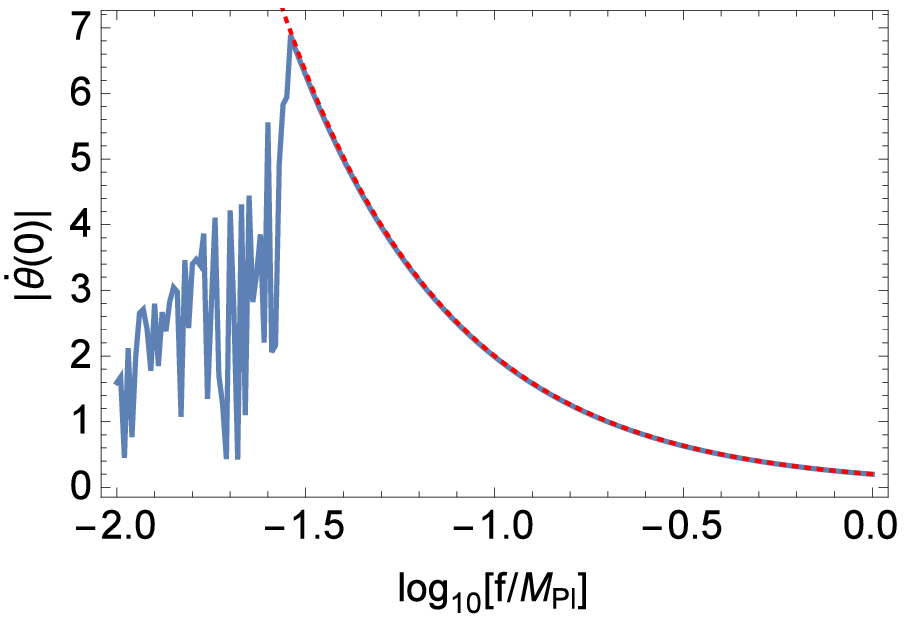}
  \end{center}
 \end{minipage}
 \begin{minipage}{0.5\hsize}
  \begin{center}
    \includegraphics[clip,width=7.0cm]{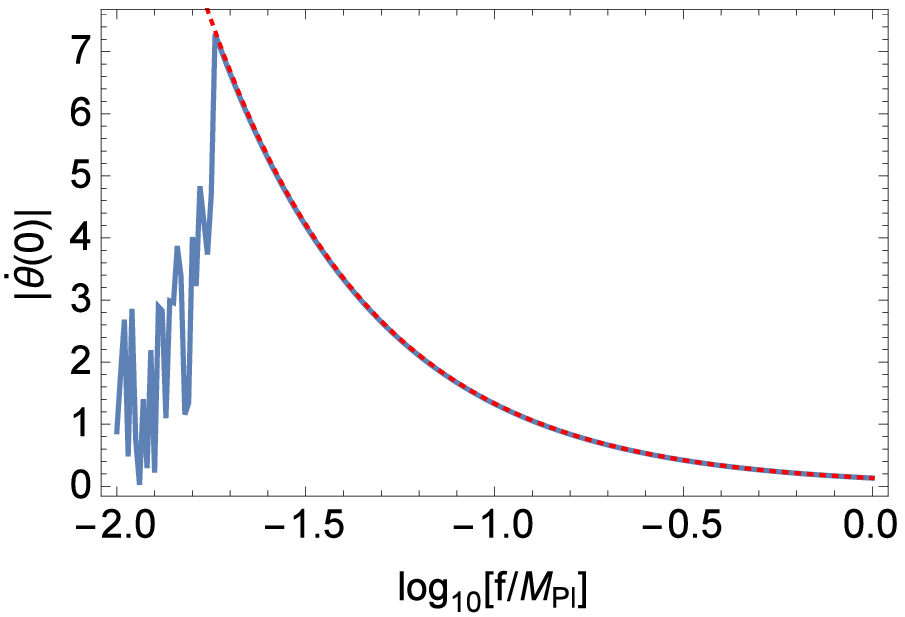}
  \end{center}
 \end{minipage}
    \caption{The velocity of $\theta$ at the end of inflation as a function of decay constant of 
    axion-inflaton $f$ in the left panel with $n=4$ and right panel with $n=6$. 
    In both panels, we set the parameters: $c=9/4$ and $|a_H|=1$. When $f\gtrsim {\cal O}(10^{-1}M_{\rm Pl})$, 
    the numerical values are well fitted with the curves, $0.2\times 10^{-\ln_{10}(f/M_{\rm Pl})}$ in the left panel 
    and $0.13\times 10^{-\ln_{10}(f/M_{\rm Pl})}$ in the right panel as expected in the 
    analytical formula~(\ref{eq:anavelo}).}
         \label{fig:infn46}
\end{figure}

By setting the parameters $c=9/4$ and $|a_H|=1$, we plot the 
velocity of $\theta$ at the end of inflation as a function of axion decay constant~$f$ 
in Fig.~\ref{fig:infn46}.
This figure shows that $\theta$ has a sizable velocity in the unit of inflaton 
mass $m_{\rm inf}\simeq 1.7\times10^{13}$GeV at the end of inflation for 
both $n=4$ and $n=6$ cases. 
Here, we mainly work with the value of decay constant within the range $10^{-2}\leq f \leq 1$ in the $M_{\rm Pl}=1$ unit, 
since the typical string theory predicts such a rage of decay constant~\cite{Choi:1985je,Svrcek:2006yi,Banks:2003sx}. 
The trans-Planckian axion decay constant is required to consider the alignment mechanism~\cite{Kim:2004rp}. 
From the numerical simulation, the velocity at the end of inflation inversely decreases with the decay constant 
of axion-inflaton $f$ when $f\gtrsim {\cal O}(10^{-1}M_{\rm Pl})$, 
which is well consistent with the analytical estimation in Eq.~(\ref{eq:anavelo}). 
Finally, we comment on the isocurvature perturbation of $\theta$ during the inflation. 
Since the phase direction of the AD field has the mass of the Hubble-scale during the inflation, 
its isocurvature perturbation can be suppressed~\cite{Enqvist:1998pf,Enqvist:1999hv,Kawasaki:2001in,Kasuya:2008xp}.

Although the previous result is concentrated on the axion monodromy inflation with quadratic form so far, 
this AD mechanism is applicable to other string axion inflation models through the axion-dependent higher-dimensional 
operators. Especially, the aligned natural inflation is discussed in Appendix.~\ref{app}.

\section{Conclusion}
\label{sec:con}
We have studied the AD baryogenesis on an axion background which is suggested by 
higher-dimensional theories and string theories. 
In contrast to the conventional AD scenario, the phase of $A$-terms and non-renormalizable operators 
are generically axion dependent 
as shown in perturbative computations and non-perturbative calculations due to the instanton effects 
in superstring theory. 
By taking into account such an axionic coupling, it is found that the axion oscillation induces the 
oscillation of the phase direction of the AD field and 
consequently the baryon asymmetry of the Universe is sufficiently produced just after the axion oscillation. 
In particular, we estimate the baryon asymmetry analytically and numerically in two scenarios where 
the power of non-renormalizable operators is $4$ in Sec.~\ref{sec:2_3_1} and $6$ in Sec.~\ref{sec:2_3_2}. 

In this paper, we focus on the situation that the axion plays a role of inflaton field such as 
the (modulated) natural inflation~\cite{Freese:1990rb,Abe:2014pwa,Czerny:2014wza,Abe:2014xja,Kappl:2015esy,Choi:2015aem}, axion monodromy inflation~\cite{Silverstein:2008sg,McAllister:2008hb}, and $F$-term axion monodromy inflation~\cite{Marchesano:2014mla,Hebecker:2015rya,Kobayashi:2015aaa}. 
In Appendix~\ref{app}, the potential of the AD field is derived in the case of 
aligned natural inflation by including the relevant axionic couplings. 
Our mechanism is applicable to the inflation models with low-reheating temperature as recently discussed 
in Refs.~\cite{Kadota:2016jlw,Parameswaran:2016qqq,Kobayashi:2016vcx}. 
Furthermore, our scenario is also applicable to  other string axion and axion-like particles, which are not 
the inflaton, but  oscillate around the minimum before the oscillating time of the AD field.

\section*{Acknowledgments}
The authors would like to thank C.~S.~Shin, K.~Sinha, N.~Takeda and F.~Takahashi for useful discussions and comments. 
 T.K. is supported in part by
the Grant-in-Aid for Scientific Research No.~26247042 from the Ministry of Education,
Culture, Sports, Science and Technology  in Japan.

\appendix

\section{Explicit models}
\label{app}
In the following, we derive the potential of the AD field by setting a specific axion inflation scenario. 
In particular, we discuss the dynamics of the AD field in the case of 
aligned natural inflation based on a string-inspired model along the line of Ref.~\cite{Kappl:2015pxa}, 
where the low-scale SUSY-breaking and high-scale inflation are realized simultaneously. 
(See also, Ref.~\cite{Abe:2014vca}, in which the realization of low-scale SUSY-breaking and high-scale inflation 
is discussed in the framework of five-dimensional supergravity.)

\subsection{Inflaton potential}
\label{app:inf}
To achieve the aligned natural inflation, 
we consider the following K\"ahler potential and superpotential\footnote{In this appendix, we employ the 
reduced Planck unit $M_{\rm Pl}=2.4\times 10^{18}\,{\rm GeV}=1$.},
\begin{align}
K&=\sum_{i=1}^2\biggl[-\ln (T_i+\bar{T}_i) +Z_i(T+\bar{T})|X_i|^2\biggl],
\nonumber\\
W&=\sum_{i=1}^2 m_i^2 X_i \left( e^{-a_iT_1-b_iT_2}-\lambda_i\right),
\label{eq:appstab}
\end{align}
where $X_i$ are the matter fields with K\"ahler metrics $Z_i$ and $T_i$ correspond to 
certain moduli fields such as K\"ahler moduli in type IIB string setup. 
Such a superpotential can be induced by the world-sheet instantons and/or 
D-brane instantons and the matter fields play a role of stabilizer fields. 
Note that the parameters $m_i, \lambda_i$ can be chosen as the real constants 
under the field redefinitions of $X_{1,2}$ and $T_{1,2}$. 

From the scalar potential constructed by the K\"ahler potential and superpotential 
in Eq.~(\ref{eq:appstab}), the fields are stabilized at the supersymmetric Minkowski minimum:
\begin{align}
&X_1=X_2=0,
\nonumber\\
&T_{1,0}=\frac{b_2\ln \lambda_1-b_1\ln \lambda_2}{a_2b_1-a_1b_2},
\nonumber\\
&T_{2,0}=\frac{a_2\ln \lambda_1-a_1\ln \lambda_2}{b_2a_1-b_1a_2},
\end{align}
at which the superpotential vanishes. 
Thus, this potential is irrelevant to the SUSY-breaking 
and at the moment, we concentrate on these moduli and stabilizer fields. 

As discussed in Ref.~\cite{Kappl:2015pxa}, for simplicity, we assume that $\lambda_2m_2^2 \gg \lambda_1m_1^2$ 
such that  one linear combination of moduli fields $a_2T_1+b_2T_2$ is 
heavier than its orthogonal combination. 
By redefining the moduli fields as
\begin{align}
T_1=T_{1,0}+b_2T +a_2\tilde{T},\nonumber\\
T_2=T_{2,0}-a_2T +b_2\tilde{T},
\end{align}
$\tilde{T}$ can be stabilized at $\tilde{T}=0$ because of the large supersymmetric mass. 
On the other hand, $T$ becomes lighter than $\tilde{T}$ 
and its scalar potential is extracted on the hypersurface $\tilde{T}=X_1=X_2=0$,
\begin{align}
V=\frac{\lambda_1^2m_1^4e^{-\delta t}\bigl[ \cosh (\delta t)-\cos (\delta \psi)\bigl]}
{2(T_{1,0}+b_2t)(T_{2,0}-a_2t)},
\end{align}
where $T=t+i \psi$ and $\delta =a_1b_2-a_2b_1$. 
The saxion $t$ becomes heavier than the axion $\psi$ at the minimum between 
two poles $t=-T_{1,0}/b_2$ and $t=T_{2,0}/a_2$ and consequently 
axion can be identified with the axion-inflaton. 
By canonically normalizing the axion $\varphi \equiv \sqrt{2b_2^2K_{T_1\bar{T}_1}-2a_2^2K_{T_2\bar{T}_2}}\psi$, 
the potential of axion-inflaton is the form of natural inflation:
\begin{align}
V_{\rm inf}=A \biggl[ 1-\cos \left(\frac{\varphi}{f}\right)\biggl],
\end{align}
where $A$ is the constant determined by the saxion and heavy modulus field, 
and the decay constant of axion-inflaton $f= \sqrt{2b_2^2K_{T_1\bar{T}_1}-2a_2^2K_{T_2\bar{T}_2}}/\delta$ is 
enhanced to be the trans-Planckian value by tuning $\delta \ll 1$ through the alignment mechanism~\cite{Kim:2004rp}. 
Finally we comment on the dynamics of saxion field. 
After the inflation, the saxion also oscillates around the minimum in a fashion similar to the axion-inflaton 
as discussed in Ref.~\cite{Abe:2014vca}, and both fields decay into the particles in the standard model. 
When both fields couple with the gauge bosons through the gauge kinetic function for the case of 
K\"ahler modulus $T$, they mainly decay into them at the same decay time.

\subsection{Potential for the AD field}
In the following, we derive the relevant potential of the AD field in Eq.~(\ref{eq:po}) specifying 
the coupling between the AD field and inflaton field step by step. 
\medskip

\noindent
$\bullet$ Negative Hubble-induced mass  

\medskip
First, we study the negative Hubble-induced mass of the AD field. 
To achieve the AD baryogenesis on an axion background, the AD field should have the negative Hubble-induced 
mass, otherwise the radial direction of the AD field does not obtain the nonvanishing VEV 
during the inflation and the baryon asymmetry has been never produced. 
As the potential of the AD field, we add the following K\"ahler potential to Eq.~(\ref{eq:appstab}),
\begin{align}
K=Z(T+\bar{T})|\Phi|^2 +\sum_{i=1}^2c_i(T+\bar{T})|X_i|^2|\Phi|^2,
\end{align}
where $Z(T+\bar{T})$ is the K\"ahler metric of the AD field 
and $c_i(T+\bar{T})$ denote the positive moduli-dependent constants. 
During the inflation, the mass squared of the AD field is estimated as
\begin{align}
V&\simeq e^K K^{X_1\bar{X}_1}|D_{X_1}W|^2 =e^K \frac{|D_{X_1}W|^2}{K_{X_1\bar{X}_1}}
\nonumber\\
&\simeq V_{\rm inf}+\left(Z-\frac{c_1}{Z_1}\right)V_{\rm inf}|\Phi|^2 
+{\cal O}(|\Phi|^4),
\end{align}
from which the AD field has the negative Hubble-induced mass when 
$Z<c_1/Z_1$. 
Although certain constraints are pointed out to implement both the inflation and 
AD baryogenesis in Refs.~\cite{Casas:1997uk,Dutta:2010sg,Marsh:2011ud,Dutta:2012mw}, 
it is only applicable to the model where the Hubble-induced mass is generated from the 
nonvanishing $F$-term of inflaton field. 
However, in our model, the negative Hubble-induced mass is achieved by the 
nonvanishing $F$-term of the stabilizer field in contrast 
to that of inflaton field as discussed in Refs.~\cite{Casas:1997uk,Dutta:2010sg,Marsh:2011ud,Dutta:2012mw}.

\medskip

\noindent
$\bullet$ Hubble-induced $A$-term  

\medskip
Next, we derive the Hubble-induced $A$-term by adding the following superpotential of the AD field,
\begin{align}
W_{\rm AD}=\left(\lambda +\lambda^\prime X_1+\lambda^{''}X \right) \Phi^n,
\label{eq:WADapp}
\end{align}
where $\lambda,\lambda^\prime,\lambda^{''}$ represent the axion dependent functions as outlined in Sec.~\ref{sec:2_2}; 
X is the SUSY-breaking field as detailed below. 
Such non-renormalizable operators are expected to be induced by the D-brane instanton effects 
and they depend on the stabilizer field to derive the 
Hubble-induced $A$-term. 
From the above coupling, the Hubble-induced $A$-term is estimated as~\cite{Kaplunovsky:1993rd,Choi:2005ge},
\begin{align}
a_H\simeq -H^{-1}F^{X_1}\partial_{X_1}\ln \left(\lambda+\lambda^\prime X_1\right)
\simeq -H^{-1}F^{X_1} \frac{\lambda^\prime}{\lambda}\sim \frac{\lambda^\prime}{\lambda},
\end{align}
which is the same order as Eq.~(\ref{eq:po}) presented in Sec.~$2$.\footnote{Now, we 
define $a_H$ as shown in Eq.~(\ref{eq:po}) in contrast to Refs.~\cite{Kaplunovsky:1993rd,Choi:2005ge}.}
Note that the phase of Hubble-induced $A$-term is determined by those of $\lambda^\prime$ and $\lambda$.

\medskip

\noindent
$\bullet$ Soft scalar mass  

\medskip
We turn to the soft terms of the AD field after the inflation. 
The discussed inflation sector does not break the SUSY at the minimum. 
To realize the SUSY-breaking at the vacuum, let us consider the SUSY-breaking 
sector with the following K\"ahler potential and superpotential,
\begin{align}
W&=w+\mu X,
\nonumber\\
K&=Z^{(1)}|X|^2 -\frac{Z^{(2)}}{\Lambda_\ast^2}|X|^4 +Z^{(3)}|X|^2|\Phi|^2,
\label{eq:SUSYbpo}
\end{align}
where $Z^{(i)}$ with $i=1,2,3$ are the moduli dependent functions; $w$ is the 
complex constant; $\mu$ can be chosen as real by the phase rotation of $X$; 
$X$ denotes 
the SUSY-breaking field; and the four-point coupling of $X$ in the K\"ahler potential is expected to 
appear from loop-corrections at the dynamical scale $\Lambda_\ast$~\cite{Kallosh:2006dv,Kitano:2006wz}. 
The SUSY-breaking field has then the nonvanishing (real) $F$-term $F^X\simeq -e^{K/2}\frac{\mu}{Z^{(1)}}$ 
at the minimum $X\ll 1$. 
During and after the inflation, $X$ is also settled into the minimum $X\ll 1$, 
where the tiny expectation value of $X$ is achieved by the tiny $\Lambda$ in the $M_{\rm Pl}$ unit. 
Thus, one can neglect the oscillation of $X$ during and after the inflation. 

Furthermore, when the four-point interaction between the AD field and $X$ exists in the 
K\"ahler potential, the canonically normalized AD field has the soft scalar mass at the 
tree-level~\cite{Kaplunovsky:1993rd,Choi:2005ge},
\begin{align}
m_{\Phi}^2\simeq m_{3/2}^2 -F^XF^{\bar{X}}\frac{Z^{(3)}}{Z},
\end{align}
which is close to the gravitino mass. Even if the AD field does not couple with the SUSY-breaking field, the soft scalar 
mass is induced by the anomaly mediation~\cite{Randall:1998uk,Giudice:1998xp}. 
Note that the soft scalar mass of the AD field should be much smaller than the Hubble-scale during the inflation, $m_\Phi, m_{3/2} <H$, 
otherwise the inflation mechanism is spoiled by the SUSY-breaking effects. 
Thus, in this model, the high-scale inflation is compatible with the low-scale SUSY-breaking. 
The AD field oscillates after the inflation at the time $t_{\rm osc}\simeq m_\Phi^{-1}\simeq m_{3/2}^{-1}$. 
\medskip

\noindent
$\bullet$ Soft SUSY-breaking $A$-term

\medskip
Next, we derive the soft SUSY-breaking $A$-term at the minimum. 
When the non-renormalizable operator of the AD field 
is a function of the SUSY-breaking field $X$ as shown in Eq.~(\ref{eq:WADapp}), 
the soft $A$-term is calculated as~\cite{Kaplunovsky:1993rd,Choi:2005ge},
\begin{align}
a_m\simeq -m_{3/2}^{-1}F^X\partial_X\ln \left( \lambda +\lambda^{''}X\right)
\simeq -m_{3/2}^{-1}F^X \frac{\lambda^{''}}{\lambda}\sim \frac{\lambda^{''}}{\lambda},
\end{align}
which is the same order as Eq.~(\ref{eq:po}) presented in Sec.~$2$.\footnote{Now, we 
define $a_m$ as shown in Eq.~(\ref{eq:po}) in contrast to Refs.~\cite{Kaplunovsky:1993rd,Choi:2005ge}.} 
Since the vacuum expectation value of $F^X$ is real, the phase of the soft $A$-term 
is determined by those of $\lambda^{''}$ and $\lambda$. 
Thus, the phase of the Hubble-induced $A$-term can be taken as the same with that of soft $A$-term when 
${\rm arg}(\lambda^{''})={\rm arg}(\lambda^\prime)$, i.e., ${\rm arg}(a_H)={\rm arg}(a_m)$ 
in the notation of Sec.~$2$. 
By contrast, when $\lambda$ is independent of $X$, the one-loop anomaly mediation 
contributes to the soft $A$-term. However, in our discussion, one can obtain the sizable 
baryon asymmetry even without the soft $A$-term. 
\medskip

\noindent
$\bullet$ $F$-term of the AD field

\medskip
Finally, we comment on the $F$-term of the AD field. 
From Eq.~(\ref{eq:WADapp}), the dominant $F$-term of the AD field 
is written as
\begin{align}
e^KK^{\Phi \bar{\Phi}}|W_\Phi|^2\simeq e^K \frac{n|\lambda|^2|}{Z}|\Phi|^{2(n-1)}.
\end{align}
Following the above procedure, we acquire the total scalar potential of the AD field 
as discussed in Eq.~(\ref{eq:po}). This potential is valid unless the SUSY-breaking scale 
is lower than the inflation scale. 
By contrast, when the $\lambda$ is zero, the $F$-term of the AD field is dominated by
\begin{align}
e^KK^{X_1 \bar{X}_1}|W_{X_1}|^2 +
e^KK^{X \bar{X}}|W_{X}|^2 
\simeq e^K n\left(\frac{|\lambda^\prime|^2}{Z_1}+\frac{|\lambda^{''}|^2}{Z^{(1)}}\right)|\Phi|^{2n}+\cdots,
\end{align}
where the dot is the irrelevant part for its $F$-term.


\end{document}